\documentclass[a4paper,
               biblatex,     
               keeplastbox,   
               ]{jacow}
\usepackage{pdfpages,multirow,ragged2e} %
\makeatletter%
	\ifboolexpr{bool{xetex}}
	 {\renewcommand{\Gin@extensions}{.pdf,%
	                    .png,.jpg,.bmp,.pict,.tif,.psd,.mac,.sga,.tga,.gif,%
	                    .eps,.ps,%
	                    }}{}
\makeatother

\ifboolexpr{bool{xetex} or bool{luatex}} 
 {}                                      
 {\usepackage[utf8]{inputenc}}           

\usepackage[USenglish]{babel}

\ifboolexpr{bool{jacowbiblatex}}%
 {%
  \addbibresource{THPR37.bib}
 }{}
\newcommand{\code}[1]{\texttt{#1}}
\newcommand{\fig}[1]{Fig.~\ref{#1}}

\newcommand{\eg}{e.g.}

\begin{document}

\title{Towards Unlocking Insights from Logbooks Using AI}

\author{A. Sulc, G. Hartmann, HZB, Berlin, DE\\
J. Maldonado, BNL, New York, USA\\
V. Kain, F. Rehm, CERN, Meyrin, CH\\
A. Eichler, J. Kaiser, T. Wilksen, F. Mayet, R.Kammering, H. Tuennermann, DESY, Hamburg, DE \\
J. St. John, H. Hoschouer, K. J. Hazelwood, FNAL, Batavia, IL, USA\\
T. Hellert, LBNL, Berkeley, CA, USA\\
D. Ratner, W.-L. Hu, A. Bien, SLAC, Menlo Park, CA, USA
}
	
\maketitle
\begin{abstract}
Electronic logbooks contain valuable information about activities and events concerning their associated particle accelerator facilities. However, the highly technical nature of logbook entries can hinder their usability and automation. As natural language processing (NLP) continues advancing, it offers opportunities to address various challenges that logbooks present. This work explores jointly testing a tailored Retrieval Augmented Generation (RAG) model for enhancing the usability of particle accelerator logbooks at institutes like DESY, BESSY, Fermilab, BNL, SLAC, LBNL, and CERN. The RAG model uses a corpus built on logbook contributions and aims to unlock insights from these logbooks by leveraging retrieval over facility datasets, including discussion about potential multimodal sources. Our goals are to increase the FAIR-ness (findability, accessibility, interoperability, and reusability) of logbooks by exploiting their information content to streamline everyday use, enable macro-analysis for root cause analysis, and facilitate problem-solving automation.
\end{abstract}

\section{Introduction}

While large language models (LLMs) have aided operators~\cite{mehta2023towards,mayet2024building, mayet2024gaia}, particle accelerator electronic logbooks (eLogs) remain underutilized due to their non-standard nature, employing highly technical language with content hidden in metadata, and privacy concerns. This challenges traditional LLMs in processing eLogs' valuable operational insights. 

In this work~\cite{techpubsnumbers}, we aim to leverage Retrieval Augmented Generation (RAG) models, tailored for eLogs, to enhance the usability and make the first steps towards automation capabilities of these highly technical data sources across participating facilities and summarize their current progress.

\section{Related Work}

In~\cite{mehta2023towards} the authors leverage notes from the fusion experiments DIII-D and Alcator C-Mod to develop a prototype "copilot" using RAG and off-the-shelf LLMs like \code{llama2-70b}. 
Their work demonstrates advantages in semantic search of experiments and assisting with device-specific operations compared to base language models and keyword search, mostly thanks to RAG combined with powerful \code{llama-2-70b}.
Multiple particle accelerator facilities are currently working on their solution for how to lift their language sources. 
We are adopting a similar approach as~\cite{mehta2023towards}, although it turned out that many eLogs have some specific challenges discussed later.

CERN is developing AccGPT~\cite{rehm2024accgpt}, an ongoing project to integrate AI assistants into particle accelerator operations, including control room assistance, coding aid, documentation, and knowledge retrieval enhancement while streamlining FAQs. A prototype capable of retrieving knowledge from various CERN internal documentation is already in production.

At DESY, Mayet developed a RAG~\cite{mayet2024building, mayet2024gaia} using \code{mixtral-8x7b-instruct-v0.1} for the SINBAD-ARES facility. 
The system further enhances user experience with the eLog with a chain of thought~\cite{wei2022chain} and ReAct prompting~\cite{yao2022react}. 
The system uses various data sources including eLog, DESY DOOCS control system, machine-specific documentation, and even facility chat groups.

The work of~\cite{maldonado2023tumbcmo15} showed a logbook analysis that benefits from multimodal sources by enhancing the eLog with text extracted from images via OCR. 
The main novelty of this work is the combined use of images and topic models to tailor and enhance the eLog system at BNL.

PACuna~\cite{sulc2023pacuna} mines accelerator data to generate Q\&A pairs for fine-tuning LMs, yielding tailored models for accelerator queries general LLMs can't handle, enabling facility-specific intelligent assistants.

\section{RAG}

The RAG pipeline is an approach that combines retrieving relevant documents from a large corpus with language generation models. 

First, given an input query $q$, an ordered list of $N$ relevant documents $\mathcal{D} = (d_1,\dots d_N)$ are retrieved using dense vector indexing. In the second step, $\mathcal{D}$ and $q$ are then passed to a generator (an LLM), which generates the final answer $a$ with $q$ and the retrieved documents $\mathcal{D}$ plugged into a predefined query template, 

\begin{equation}
    q\rightarrow\underset{\mathcal{D}}{\underbrace{\text{retrieve}(q)}}\rightarrow\underset{a}{\underbrace{\text{generate}(q,\mathcal{D})}}\rightarrow a.
\end{equation}

This allows language models to hallucinate less and utilize real-world knowledge present in the retrieved documents $\mathcal{D}$ instead of internal (trained) model parameters. 

\subsection{Document Retrieval - $\text{retrieve}(q)$}

Embedding is crucial for retrieval, but off-the-shelf models lack domain knowledge. We fine-tuned on domain data for SimCSE~\cite{gao2021simcse} but struggled with longer texts even after fine-tuning \code{sci-bert-\{cased, uncased\}}. The cased version performed better, but \code{all-MiniLM-L6-v2} outperformed while balancing size and accuracy.

Re-ranking~\cite{shakir2024boost} can improve performance by filtering and re-ranking based on relevance scores from a specialized model. Only top-ranked documents are passed to the generator, focusing on pertinent information.

\section{Current Status}

This section will provide a concise overview of the existing eLog technology and outline the plans.

\subsection{ALS}

%
At the ALS, we've enhanced semantic logbook searches due to acronym/jargon use. Initial word2vec~\cite{mikolov2013efficient} attempts were unsatisfactory, prompting robust solutions. We're refining an embedding model and integrating \code{all-MiniLM-L6-v2} into the RAG pipeline with \code{Mistral-7B-Instruct-v0.2}.

Additionally, we've started recording weekly Operations Critic Meetings for RAG pipeline inclusion. Transcribing these faces hurdles due to poor Zoom audio quality, multiple in-room participants, and prevalent acronym/colloquial language use.


\subsection{BNL}

\subsubsection{Techniques}
The BNL eLog system used at the Collider-Accelerator Department (C-AD) allows users to customize logbook settings, including the specification of favorite logbooks. 
Our goal was to use ML techniques to further personalize configurations and provide users with entries that match their specific interests. 
This work will benefit users who troubleshoot systems with a need for quick and accurate information to avoid delays in operational time.

We utilized Gensim~\cite{rehurek2011gensim} tools and the Doc2Vec~\cite{le2014distributed} model to augment the search feature. 
The eLog data is stored in a MySQL database and imported using custom tools into a Pandas dataframe. 
The content of the eLog entries was transformed into vectors of words after tokenization, lemmatization, and removing punctuation. 
Initial tests showed the model was able to find similarities between common words like \code{polarization}, \code{beta}, \code{bunch}, \code{coherence}, and \code{emittances}. 
Classification techniques were explored to identify major topics in entries. These topics are useful as tags for grouping entries for users.

\subsubsection{Web Application} 
After testing, this search feature will be added to eLog. A temporary web demo~\cite{maldonado2024improving} lets users test it by entering words/phrases to get similar entries. The text is processed through pretrained models, displaying search results with date, author, log name, similarity, and link.

\subsubsection{Future Work} 
For personalization, we aim to develop a reinforcement learning model using user feedback rewards. User feedback will also improve the workflow for eLog system implementation.

\subsection{CERN}
CERN is currently investing in building a pilot service with the name AccGPT~\cite{rehm2024accgpt} for hosting LLMs either fine-tuned or out-of-the-box with RAG based on CERN internal knowledge. For the time being, AccGPT will not incorporate logbooks due to concerns about their potentially lower data quality. With the idea that in the future, LLMs and AI, in general, could be used to fill next-generation logbooks and provide accelerator statistics. The team at CERN believes that this will make knowledge from logbooks and other accelerator data more exploitable. In the middle of 2024, AccGPT will go through its first accuracy tests with domain experts. At the same time benchmarking Q\&A datasets are being curated.

\subsection{DESY \& BESSY}

DESY and BESSY are developing new eLog simultaneously. 
DESY has already shown success in implementing RAGs at SINBAD-ARES~\cite{mayet2024building, mayet2024gaia}. 
The implementation combines multiple sources such as the logbook itself, data obtained from the interaction with the control system, operations meeting slides, Mattermost chats, and the accelerator lattice (accelerator layout).
\vspace{-1em}
\subsubsection{Embedding, Vectorstore and Generator}
For DESY European XFEL, we experimented with a fine-tuned embedding model as mentioned earlier, but smaller generic models like \code{all-MiniLM-L6-v2}~\cite{reimers-2019-sentence-bert}, trained on supervised data still perform better. We also experimented with recent high-quality embedding models~\code{mxbai-embed-large-v1}~\cite{emb2024mxbai,li2023angle} and \code{nomic-embed-text-v1}~\cite{nussbaum2024nomic}.

We use FAISS vectorstore~\cite{johnson2019billion} due to its simplicity and \code{Mistral-7B-Instruct-v0.2} as a generator.

\vspace{-1em}
\subsubsection{Metadata}
We increased the verbosity of eLog entries by storing control system metadata from our control system when screenshots are made, see~\fig{fig:desy-metadata}.
\vspace{-1em}
\subsubsection{Future Work}
Recall is a bottleneck. Enhancing training data/benchmarks is needed.
The next steps will lead to automating/augmenting eLog entries by processing screenshots (\cite{zhang2024tinychart}), using achiever data, or enhancing from the machine/control system state. Use LLMs to enhance posts with specialized generators~\cite{sulc2023pacuna} or RAFT~\cite{zhang2024raft} (better RAG models).

\iftrue
\begin{figure}[!hb]
    \centering
    \begin{minipage}{0.45\linewidth}
            \includegraphics[width=1.0\linewidth]{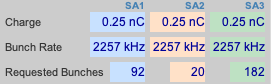}
    \end{minipage}    
    \begin{minipage}{0.45\linewidth}
        \resizebox{1.1\linewidth}{!}{
        \begin{tabular}{l}
            \code{Value: .../TORA.25.I1/CHARGE.SA3 0.25 nC}\\
            \code{Value: .../TORC.3098.T4D/NUMBEROFBUNCHES.SA3 182}\\
            \code{Value: .../TORA.25.I1/CHARGE.SA2 0.25 nC}\\
            \code{Value: .../TORC.3181.T5D/NUMBEROFBUNCHES.SA2 20}\\
            \code{Value: .../TORA.25.I1/CHARGE.SA1 0.25 nC}\\
            \code{Value: .../TORC.3098.T4D/NUMBEROFBUNCHES.SA1 92}
        \end{tabular}}
    \end{minipage}
    \caption{Stored metadata from the control system.}
    \label{fig:desy-metadata}
\end{figure}
\fi

We're exploring approaches like topic modeling~\cite{grootendorst2022bertopic}, root cause analysis~\cite{roy2024exploring,ahmed2023recommending}, and using large language models for autonomous accelerator tuning~\cite{kaiser2024large}.

\subsection{Fermilab}

\subsubsection{ADEL}
The Fermilab Accelerator Directorate Electronic Logbook (ADEL)~\cite{hazelwood_wao_2023} is the primary record of accelerator operations at Fermilab. Everything from machine failures to study notes is recorded and categorized there. Since its creation in 2013, ADEL has recorded nearly one million user-generated entries, comments, and files. ADEL data may be filtered by ID, date, text, subject, log, category, and user. All data is saved in a relational database and a rich HTTP client and program API exists to read or extract data.

\vspace{-1em}
\subsubsection{Past Work}
There was some effort in the past to introduce AI/ML into ADEL. The first attempt focused on using Optical Character Recognition (OCR) to capture text from the many thousands of controls system screenshots that are often posted in the eLog during tuning and failure diagnosis.
The other use of AI/ML for ADEL was to categorize images to improve search~\cite{njekeu_noice_2020}. 
There was moderate success training a model to classify images by which controls system GUI they originated from or if the image was of a photo or a screenshot. There were plans to run the trained image classifier model on the eLog data and save the categorization to a new database table field, but lack of resources and time hampered the work and it has not yet been implemented.

\vspace{-1.0em}
\subsubsection{Current Work}
Fermilab has developed a semantic search prototype capable of searching ADEL quickly and retrieving relevant results. Accelerator Operators frequently use the Elog when troubleshooting everyday issues and have found the semantic search results to be more useful and relevant than lexical search results.

\vspace{-1.0em}
\paragraph{Embedding, Vectorstore, Re-Ranking and Recommendations}
For embedding entries, \code{all-mpnet-base-v2}\cite{reimers-2019-sentence-bert} proved to be very powerful even compared to the unsupervised SimCSE fine-tuned model which is expected to be biased towards shorter log entries.

Qdrant\cite{Qdrant} was used as the vector store. It is an open-source, locally hosted solution that allows for storing vectors alongside associated metadata, providing useful filtering options such as date, topic, and so on.

Additionally, a re-ranking function was implemented using cross-encoding models. 
Tested models for the re-ranking include \code{mxbai-rerank-base-v1}\cite{shakir2024boost} and \code{ms-marco-MiniLM-L-12-v2}\cite{reimers-2020-Curse_Dense_Retrieval}. 
The re-ranking produces very relevant results, even without filtering, allowing retrieval of notes that may have been mislabeled. 
Qdrant also supports a recommendation feature, allowing it to push the query closer to relevant points.

\vspace{-1.25em}
\subsubsection{Future work:}
The Fermilab team plans to fully integrate the search function with the Elog (ADEL) soon and also to incorporate image-similarity search.

\subsection{SLAC}
SLAC is investigating both text generation and natural-language queries (NLQ). Within text generation, LLMs can assist in generating routine reports, \eg{} shift summaries, as well as assist in creating entries for future queries, \eg{} by identifying ambiguous terms or suggesting keywords.

NLQ will help filter ambiguous terms (\eg{} \code{beam energy}), select related terms (\eg{} multiple names for an RF station), or identify historical entries that are similar to a target entry. 
More ambitiously, queries could perform calculations, such as computing variance in repeated measurements or total time spent on a task category. 
\vspace{-1.0em}
\subsubsection{Datasets}

Natural language datasets comprise eLogs for major accelerator facilities for both physicists and operators. Operator manuals (\eg{} the `wiki'), technical documents and user-side eLogs provide additional context.
Databases of GUI measurements and auto-archiving of diagnostic measurements, accelerator setpoints, and high-level analysis, can be linked to logbook entries through time stamps. 
Concerns exist that personally identifiable information (PII) in logbooks may require scrubbing or access limitation. As a precautionary measure, we have established a data processing workflow and are applying for an Institutional Review Board (IRB) review to ensure that our research complies with regulations and protects PII. User-side eLogs require permission from authors and are excluded from early work.

\vspace{-1.0em}
\subsubsection{Initial work}
%
%
The first application of LLMs at SLAC augments logbook interpretation using an internal operations Mediawiki. Whitelisted Wiki articles support a RAG pipeline generating operations-backed context. A 'cleaning' stage prompts an LLM to read an article, generate Q\&A, and prompt staff for validation. The process updates the wiki and produces question-answer pairs for fine-tuning the LLM.

\section{Conclusion}
This collaborative work demonstrated initial progress in implementing RAG models to unlock insights from particle accelerator eLogs across multiple institutes with various logbooks. 
While challenges with a technical language remain, which cause issues in document retrieval, the findings highlight the potential of leveraging AI to enhance the usability and accessibility of eLog data. 
Further research building on this foundation is crucial for realizing the full impact of NLP and the use of metadata.

\section{Acknowledgement} 
We acknowledge DESY (Hamburg, Germany) and HZB (Berlin, Germany), a member of the Helmholtz Association HGF, for their support in providing resources and infrastructure. This work has in part been funded by the IVF project InternLabs-0011 (HIR3X).
This manuscript has been authored in part by Fermi Research Alliance, LLC under Contract No. DE-AC02-07CH11359 with the U.S. Department of Energy, Office of Science, Office of High Energy Physics. SLAC work is supported by the Department of Energy, Office of Science, Office of Basic Energy Sciences under contract DE-AC02-76SF00515.

\clearpage

%
%
\ifboolexpr{bool{jacowbiblatex}}%
	{\printbibliography}%

@misc{techpubsnumbers,
    howpublished = "FERMILAB-CONF-24-0237-AD"
}

@article{hazelwood_wao_2023,
    title = {The Fermilab Accelerator Division Electronic Logbook (ADEL) at 10 Years},
    author = {Hazelwood, K. J. and Finstrom, D. and McCusker-Whiting, M. and Mills, L. G.},
    abstractNote = {The current Fermilab Accelerator Division (AD) Electronic Logbook (ADEL) wasfirst released in 2013. It’s creation a decade ago represented a marked improvement over the previous electronic logbook (eLog) and brought about the first consolidated Fermilab AD eLog built upon a relational database. Over the past ten years, ADEL has performed admirably, logging nearly one million user entered entries, comments and files, undergoing a revision and adding many additional features. However, as the work of Fermilab has changed, so has theuse of ADEL, some of which it had not been designed for.},
    doi = {10.2172/2246727},
    url = {https://www.osti.gov/biblio/2246727}, 
    place = {United States},
    year = {2023},
    month = {12}
}

@article{njekeu_noice_2020,
    title = {(NOICE) Neural Optical Image Categorizer for the E-log},
    author = {Njekeu, Terence},
    abstractNote = {The Fermilab Accelerator Division Electronic logbook (E-log) is a record of all the activities and events in the Division for the past 10 years and more. The E-log search function is a valuable resource and the institutional memory of the accelerator complex. About 300,000 files are stored in the E-log, of which the vast majority are images attached to entries and comments. The visual information contained in the images is not presently searchable. The goal of team NOICE (Neural Optical Image Categorizer for the E-log) was to design a neural network able to produce label categories for these images for use by searches. The group developed a dataset and trained a convolutional neural network (CNN) with optimized hyperparameter},
    url = {https://www.osti.gov/biblio/1706139},
    place = {United States},
    year = {2020},
    month = {8}
}

@inproceedings{mehta2023towards,
title={Towards LLMs as Operational Copilots for Fusion Reactors},
author={Mehta, Viraj and Abbate, Joseph and Wang, Allen and Rothstein, Andrew and Char, Ian and Schneider, Jeff and Kolemen, Egemen and Rea, Cristina and Garnier, Darren},
booktitle={NeurIPS 2023 AI for Science Workshop},
year={2023}
}

@article{mayet2024gaia,
  title={GAIA: A General AI Assistant for Intelligent Accelerator Operations},
  author={Mayet, Frank},
  journal={arXiv preprint arXiv:2405.01359},
  year={2024}
}

@misc{mayet2024building,
title={Building an Intelligent Accelerator Operations Assistant using Advanced Prompt Engineering Techniques and a High Level Control System Toolkit},
author={Mayet, Frank},
booktitle={1st Large Language Models in Physics Symposium (LIPS)},
year={2024},
month={2},
day={21–23}
}

@misc{rehm2024accgpt,
title={{AccGPT: A Vision for AI Assistance at CERN’s Accelerator Control and Beyond}},
author={Rehm, Florian and Guijarro, Juan Manuel and Soufflet, Nathan and Kain, Verena},
booktitle={1st Large Language Models in Physics Symposium (LIPS)},
year={2024},
month={2},
day={21–23}
}

@article{gao2021simcse,
title={Simcse: Simple contrastive learning of sentence embeddings},
author={Gao, Tianyu and Yao, Xingcheng and Chen, Danqi},
journal={arXiv preprint arXiv:2104.08821},
year={2021}
}

@inproceedings{maldonado2023tumbcmo15,
author = {J. Maldonado and S.L. Clark and W. Fu and S. Nemesure},
title = {{Enhancing Electronic Logbooks Using Machine Learning}},
booktitle = {Proc. 19th Int. Conf. Accel. Large Exp. Phys. Control Syst. (ICALEPCS’23)},
eventdate = {2023-10-09/2023-10-13},
pages = {382–385},
paper = {TUMBCMO15},
language = {english},
venue = {Cape Town, South Africa},
series = {International Conference on Accelerator and Large Experimental Physics Control Systems},
number = {19},
publisher = {JACoW Publishing, Geneva, Switzerland},
month = {02},
year = {2024},
issn = {2226-0358},
isbn = {978-3-95450-238-7},
doi = {10.18429/JACoW-ICALEPCS2023-TUMBCMO15},
url = {https://jacow.org/icalepcs2023/papers/tumbcmo15.pdf}
}

@article{wei2022chain,
title={Chain-of-thought prompting elicits reasoning in large language models},
author={Wei, Jason and Wang, Xuezhi and Schuurmans, Dale and Bosma, Maarten and Xia, Fei and Chi, Ed and Le, Quoc V and Zhou, Denny and others},
journal={Advances in neural information processing systems},
volume={35},
pages={24824–24837},
year={2022}
}

@online{emb2024mxbai,
  title={Open Source Strikes Bread - New Fluffy Embeddings Model},
  author={Sean Lee, Aamir Shakir, Darius Koenig and Julius Lipp},
  year={2024},
  url={https://www.mixedbread.ai/blog/mxbai-embed-large-v1},
}

@article{li2023angle,
  title={AnglE-optimized Text Embeddings},
  author={Li, Xianming and Li, Jing},
  journal={arXiv preprint arXiv:2309.12871},
  year={2023}
}

@article{sulc2023pacuna,
  title={PACuna: Automated Fine-Tuning of Language Models for Particle Accelerators},
  author={Sulc, Antonin and Kammering, Raimund and Eichler, Annika and Wilksen, Tim},
  journal={arXiv preprint arXiv:2310.19106},
  year={2023}
}

@article{zhang2024tinychart,
  title={TinyChart: Efficient Chart Understanding with Visual Token Merging and Program-of-Thoughts Learning},
  author={Zhang, Liang and Hu, Anwen and Xu, Haiyang and Yan, Ming and Xu, Yichen and Jin, Qin and Zhang, Ji and Huang, Fei},
  journal={arXiv preprint arXiv:2404.16635},
  year={2024}
}

@article{yao2022react,
  title={React: Synergizing reasoning and acting in language models},
  author={Yao, Shunyu and Zhao, Jeffrey and Yu, Dian and Du, Nan and Shafran, Izhak and Narasimhan, Karthik and Cao, Yuan},
  journal={arXiv preprint arXiv:2210.03629},
  year={2022}
}

@misc{shakir2024boost,
title={Boost Your Search With The Crispy mixedbread Rerank Models},
author={Shakir, Aamir and Koenig, Darius and Lipp, Julius and Lee, Sean},
howpublished={\url{https://www.mixedbread.ai/blog/mxbai-rerank-v1}},
year={2024},
month={5},
day={7},
note={Accessed: 2024-05-07}
}

@article{johnson2019billion,
  title={Billion-scale similarity search with {GPUs}},
  author={Johnson, Jeff and Douze, Matthijs and J{\'e}gou, Herv{\'e}},
  journal={IEEE Transactions on Big Data},
  volume={7},
  number={3},
  pages={535--547},
  year={2019},
  publisher={IEEE}
}

@article{rehurek2011gensim,
  title={Gensim--python framework for vector space modelling},
  author={Rehurek, Radim and Sojka, Petr},
  journal={NLP Centre, Faculty of Informatics, Masaryk University, Brno, Czech Republic},
  volume={3},
  number={2},
  year={2011}
}

@inproceedings{le2014distributed,
  title={Distributed representations of sentences and documents},
  author={Le, Quoc and Mikolov, Tomas},
  booktitle={International conference on machine learning},
  pages={1188--1196},
  year={2014},
  organization={PMLR}
}

@article{zhang2024raft,
  title={Raft: Adapting language model to domain specific rag},
  author={Zhang, Tianjun and Patil, Shishir G and Jain, Naman and Shen, Sheng and Zaharia, Matei and Stoica, Ion and Gonzalez, Joseph E},
  journal={arXiv preprint arXiv:2403.10131},
  year={2024}
}

@article{grootendorst2022bertopic,
  title={BERTopic: Neural topic modeling with a class-based TF-IDF procedure},
  author={Grootendorst, Maarten},
  journal={arXiv preprint arXiv:2203.05794},
  year={2022}
}

@article{roy2024exploring,
  title={Exploring LLM-based Agents for Root Cause Analysis},
  author={Roy, Devjeet and Zhang, Xuchao and Bhave, Rashi and Bansal, Chetan and Las-Casas, Pedro and Fonseca, Rodrigo and Rajmohan, Saravan},
  journal={arXiv preprint arXiv:2403.04123},
  year={2024}
}

@inproceedings{ahmed2023recommending,
  title={Recommending root-cause and mitigation steps for cloud incidents using large language models},
  author={Ahmed, Toufique and Ghosh, Supriyo and Bansal, Chetan and Zimmermann, Thomas and Zhang, Xuchao and Rajmohan, Saravan},
  booktitle={2023 IEEE/ACM 45th International Conference on Software Engineering (ICSE)},
  pages={1737--1749},
  year={2023},
  organization={IEEE}
}

@article{kaiser2024large,
      title={Large Language Models for Human-Machine Collaborative Particle Accelerator Tuning through Natural Language}, 
      author={Jan Kaiser and Annika Eichler and Anne Lauscher},
      year={2024},
      eprint={2405.08888},
      journal={arXiv preprint arXiv:2405.08888},
      archivePrefix={arXiv},
      primaryClass={cs.CL}
}

@article{mikolov2013efficient,
  title={Efficient estimation of word representations in vector space},
  author={Mikolov, Tomas and Chen, Kai and Corrado, Greg and Dean, Jeffrey},
  journal={arXiv preprint arXiv:1301.3781},
  year={2013}
}

@online{maldonado2024improving,
title={Improving Electronic Logbook Searches Using Natural Language Processing},
author={Maldonado, Jennefer and Clark, Samuel and Nemesure, Seth},
url = {https://www.indico.kr/event/47/contributions/514/},
year={2024},
month={3},
day={7},
time={17:00}
}

@software{Qdrant,
  author = {{Qdrant}},
  title = {Qdrant Vector Database},
  url = {https://https://qdrant.tech/},
  version = {1.9.2},
  month = "3",
  year = "2024"}

@inproceedings{reimers-2019-sentence-bert,
  title = "Sentence-BERT: Sentence Embeddings using Siamese BERT-Networks",
  author = "Reimers, Nils and Gurevych, Iryna",
  booktitle = "Proceedings of the 2019 Conference on Empirical Methods in Natural Language Processing",
  month = "11",
  year = "2019",
  publisher = "Association for Computational Linguistics",
  url = "https://arxiv.org/abs/1908.10084",
}

@inproceedings{reimers-2020-Curse_Dense_Retrieval,
    title = "The Curse of Dense Low-Dimensional Information Retrieval for Large Index Sizes",
    author = "Reimers, Nils  and Gurevych, Iryna",
    booktitle = "Proceedings of the 59th Annual Meeting of the Association for Computational Linguistics and the 11th International Joint Conference on Natural Language Processing (Volume 2: Short Papers)",
    month = "8",
    year = "2021",
    address = "Online",
    publisher = "Association for Computational Linguistics",
    url = "https://arxiv.org/abs/2012.14210",
    pages = "605--611",
}

@misc{nussbaum2024nomic,
      title={Nomic Embed: Training a Reproducible Long Context Text Embedder}, 
      author={Zach Nussbaum and John X. Morris and Brandon Duderstadt and Andriy Mulyar},
      year={2024},
      eprint={2402.01613},
      archivePrefix={arXiv},
      primaryClass={cs.CL}
}
	{%
	

} 
%
%


\end{document}